\documentstyle[epsf,twocolumn,twoside,graphics,graphicx,wrapfloat]{tdc} 
\newenvironment{mtitle}{\begin{trivlist} \Large \bf
            \bf \item[]}{\vspace{0.5\baselineskip}
        \end{trivlist}}
\newenvironment{mauthor}{\begin{trivlist} \raggedright \bf
             \item[]}{\end{trivlist}}

\newcommand{\insertplot}[6]{\begin{figure}
      \hfill\hbox to 0.05in{\vbox to #5pt{\vfill
       \inputplot{#1}{#4}{#5}{#6}}\hfill}
        \hfill\vspace{-.1in}\caption{#2}\label{#3}
         \end{figure}}
\newcommand{\insertplotb}[6]{\begin{figure}[b]
      \hfill\hbox to 0.05in{\vbox to #5pt{\vfill
       \inputplot{#1}{#4}{#5}{#6}}\hfill}
        \hfill\vspace{-.1in}\caption{#2}\label{#3}
         \end{figure}}
\newcommand{\insertploth}[6]{\begin{figure}[h]
      \hfill\hbox to 0.05in{\vbox to #5pt{\vfill
       \inputplot{#1}{#4}{#5}{#6}}\hfill}
        \hfill\vspace{-.1in}\caption{#2}\label{#3}
         \end{figure}}
\newcommand{\insertplotw}[6]{\begin{figure*}
      \hfill\hbox to 0.05in{\vbox to #5pt{\vfill
       \inputplot{#1}{#4}{#5}{#6}}\hfill}
        \hfill\vspace{-.1in}\caption{#2}\label{#3}
         \end{figure*}}
\newcommand{\insertplotwt}[6]{\begin{figure*}[t]
      \hfill\hbox to 0.05in{\vbox to #5pt{\vfill
       \inputplot{#1}{#4}{#5}{#6}}\hfill}
        \hfill\vspace{-.1in}\caption{#2}\label{#3}
         \end{figure*}}
\newcommand{\insertplotwr}[8]{\begin{figure*}
      \hfill\hbox to 0.05in{\vbox to #5pt{\vfill
       \inputplotr{#1}{#4}{#6}{#7}{#8}}\hfill}
        \hfill\vspace{-.1in}\caption{#2}\label{#3}
         \end{figure*}}
\newcommand{\inputplot}[4]{
  \special{epsfile=#1 hsize=#2 vsize=#3 hoffset=-#4 voffset=#3}} 
\newcommand{\inputplotr}[5]{
  \special{epsfile=#1 hsize=#2 vsize=#3 hoffset=-#4 voffset=#5}}

\newcommand{\insertphoto}[6]{\begin{photo}
      \hfill\hbox to 0.05in{\vbox to #5pt{\vfill
       \inputplot{#1}{#4}{#5}{#6}}\hfill}
        \hfill\vspace{-.1in}\caption{#2}\label{#3}
         \end{photo}}
\newcommand{\insertphotow}[6]{\begin{photo*}
      \hfill\hbox to 0.05in{\vbox to #5pt{\vfill
       \inputplot{#1}{#4}{#5}{#6}}\hfill}
        \hfill\vspace{-.1in}\caption{#2}\label{#3}
         \end{photo*}}

\def\MARU#1{{\ooalign{\hfil#1\/\hfil\crcr
    \raise.167ex\hbox{\mathhexbox20D}}}}

\begin{document}

\twocolumn  

\begin{mtitle}
Intercomparison Study of Time and Frequency Transfer between VLBI and Other Techniques \\{\it (GPS, ETS8(TCE), TW(DPN) and DMTD)}
\end{mtitle}

\begin{mauthor}
Hiroshi Takiguchi$^1$ ({\it hiroshi.takiguchi}@{\it aut.ac.nz}), \\ 
Moritaka Kimura$^2$, Tetsuro Kondo$^2$, Atsutoshi Ishii$^3$, Hobiger Thomas$^2$, Ryuichi Ichikawa$^2$, Yasuhiro Koyama$^2$, 
Yasuhiro Takahashi$^2$, Fumimaru Nakagawa$^2$, Maho Nakamura$^2$, Ryo Tabuchi$^2$, Shigeru Tsutshiya$^2$, Shinichi Hama$^2$, 
Tadahiro Gotoh$^2$, Miho Fujieda$^2$, Masanori Aida$^2$, Tingyu Li$^2$, and Jun Amagai$^2$
 \\ \vspace{\baselineskip}

$^1${\it Institute for Radio Astronomy and Space Research, Auckland University of Technology} \\
$^2${\it Space-Time Standards Laboratory, National Institute of Information and Communications Technology} \\
$^3${\it Advance Engineering Services Co., Ltd} \\

\end{mauthor}

\setcounter{section}{0}
\setcounter{subsection}{0}
\setcounter{figure}{0}
\setcounter{table}{0}
\setcounter{footnote}{0}

\hspace{-1em}{\it Abstract}: 
We carried out the intercomparison experiments between VLBI and other techniques to show the capability of VLBI time and frequency transfer by using the current geodetic VLBI technique and facilities as the summary of the experiments that we carried out since 2007.
The results from the two different types of experiments show that the VLBI is more stable than GPS but is slightly noisier than two new two-way techniques (TW(DPN), ETS8(TCE)), and VLBI can measure the correct time difference as same as ETS8(TCE).  

\section{Introduction}
As one of the new time and frequency transfer (hereafter T$\&$F transfer) technique to compare the next highly stable frequency standards, we proposed the geodetic VLBI technique \cite{taki1}. Since 2007, to evaluate the capability of geodetic VLBI for precise T$\&$F transfer, we carried out intercomparison experiments between VLBI and GPS Carrier Phase (hereafter GPS) on the Kashima 11m and Koganei 11m baseline several times. These intercomparisons showed that the geodetic VLBI technique has the potential for precise frequency transfer \cite{taki2}, \cite{taki3}. Also, these results showed that the geodetic VLBI can measure the correct time difference \cite{taki4}.

Space-Time Standards Group of National Institute of Information and Communications Technology (NICT) which we belong to, is conducting research and developments for precise T$\&$F transfer techniques other than VLBI such as using GPS and two-way satellite time and frequency transfer (TWSTFT) at NICT Koganei Headquaters.
In 2010, we carried out the intercomparison experiment between VLBI and other techniques to show the capability of VLBI time and frequency transfer by using the current geodetic VLBI technique and facilities as the summary of the experiments.

In this paper, we describe the two intercomparison experiments from a viewpoint of the VLBI mainly.
Therefore, we leave the details of the result of other techniques to different papers.

\section{Two new TWSTFT techniques developed by NICT}
NICT developed the two new TWSTFT techniques. One is the method using a pair of Pseudo Random Noises (dual PRN, DPN) (hereafter TW(DPN)). The other one is the method using Time Comparison Equipment (TCE) on the Engineering Test Satellite VIII (ETS8) (hereafter ETS8(TCE)).
To carry out the intercomparison experiment, we installed the TW(DPN) antenna and the ETS8(TCE) ground station at Kashima Space Technology Center (KSTC, former Kashima Space Research Center) next to the VLBI antenna (Kashima 11m).
In this section, we describe the brief overview of two techniques.
Please see the reference papers for more details.

\subsection{TW(DPN)}
TW(DPN) was developed to improve the measurement precision and decrease operational cost of TWSTFT.
The precision can be improved by increasing the chip rate of the PRN.
However, this method of enhancing the precision is not feasible because the rental costs of the commercial communication satellites used for signal transfer are high.

TW(DPN) is composed of a waveform generator and an A/D converter.
By using this method, we can improve the delay measurement precision by one order of magnitude, even though the occupied bandwidth is only 400kHz, which is less than one-sixth the currently used bandwidth.
Since the transponder cost is proportional to the occupied bandwidth, we can reduce the operational cost of the TWSTFT \cite{amagai}.

\subsection{ETS8(TCE)}
ETS8 is a Japanese Geostationary Satellite, which launched in 2006.
ETS8 has missions for mobile communication experiments and for precision timing experiments using Cesium atomic clocks in space.

At the time of T$\&$F transfer, TCE transmit and receive signals to and from the ground.
As the two-way uplink and downlink transmission pathways are approximately equivalent, the effects of transmission delay in the atmosphere or those due to the motion of the satellite will be cancelled out, enabling highly precise time transfer, with anticipated precision on the order of several nanoseconds in code-phase operation and approximately less than 100 picoseconds in carrier-phase operation \cite{takahashi1}, \cite{fujieda1}, \cite{nakagawa}.

\section{Intercomparison experiments}
\subsection{Outline of the experiments}
\begin{figure*}[htbp]
\begin{center}
\includegraphics[width=0.7\textwidth]{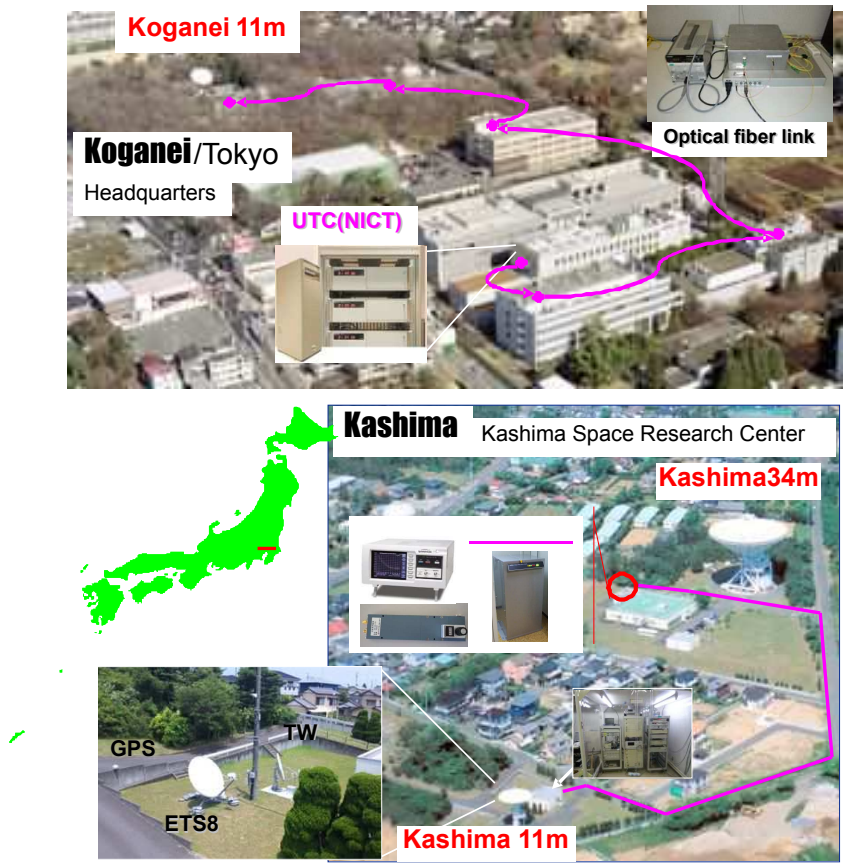}
\caption{The layout map of KSTC and Koganei Headquaters.}
\label{map}
\end{center}
\end{figure*}

Figure \ref{map} is the layout map of KSTC and Koganei Headquaters.
The baseline length of Kashima 11m - Koganei 11m is about 109 km.
In 2010, we carried out intercomparison experiments two times (August and October).
At August experiment, to evaluate long term stability of these techniques, we acquired the over 100 hours data.
At October experiment, we compared the precision of these techniques by stretching the Coaxial Phase Shifter (hereafter trombone) which was inserted in the path of the reference signal from Hydrogen maser to the Kashima 11m antenna \cite{taki4}.
The reference signal which is provided from hydrogen maser is transmitted by coaxial cable (the distance is about 300m) in KSTC.
In 2009, we installed the RF distribution system using optical fibers at Koganei Headquaters to transmit the reference signal to VLBI back end which is coherent with UTC(NICT). 
To cancel the length fluctuation of optical fibers, we adopted the feedback system using the round-trip signal. Hence, this transfer stability reached the 10$^{-16}$ level over 1000 seconds. Therefore, the reference signal (10MHz/1PPS) at Koganei station is coherent with UTC(NICT) \cite{fujieda2}.

\subsection{Geodetic VLBI using K5/VSI system}
\begin{figure}[htdp]
 \begin{center}
  \includegraphics[width=40mm]{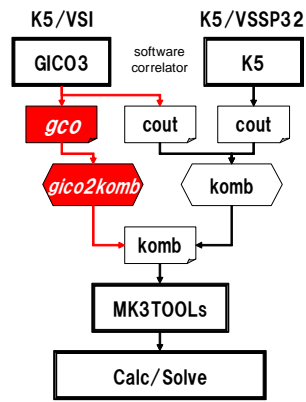}
  \caption{The flowchart from the two K5 sampling systems to the baseline analysis software.}
  \label{flow}             
\end{center}
\end{figure}

NICT are currently developing the two types of sampling system named as K5/VSSP32 (hereafter VSSP) and K5/VSI such as ADS1000 and ADS3000+ (hereafter VSI) \cite{koyama}. Also, we are developing the software correlator and data conversion utilities \cite{hobiger1} corresponding to each system. VSSP and K5 software correlator are one of the sets, and that is mainly used for the geodetic VLBI experiments in Japanese stations \cite{kondo}. VSI and GICO3 software correlator are another sets. That is mainly used for astronomical purpose.
The processing speed of the GICO3 worthy of special mention is about 10 times faster than that of the DiFX at 2k FFT points \cite{kimura}.
In order to use VSI system in the geodetic VLBI experiment, we developed the data conversion programs and carried out the experiments.
Figure \ref{flow} show the flowchart that indicate from K5 sampling system to the baseline analysis software. The gray background indicate the new programs which were developed in this time.

The setting parameters of both experiments are shown in Table \ref{vsspvsi}.
At first, we supposed VSSP was main.
Therefore, the effective bandwidth (X-band) of VSI was narrower than VSSP in spite of a wideband sampler.
In addition, because the schedule was optimized for VSSP, the scan length was longer, and the number of observation was less, than the schedule that was optimized for VSI.
As the result, the estimated delay precision of VSI was 70$\%$ with reference to VSSP.

Table \ref{baseline} shows the baseline length calculated from the two types of K5 system.
In the two time experiments, these results show the good agreement.
Thus it is concluded that using the K5/VSI and GICO3 software correlator for the geodetic VLBI experiment is not a problem.
However, the results of the clock offsets from VSSP have daily variations which were influenced from the problem of phase calibration system.
Therefore, the results of VSSP shown afterward were corrected using VSI data.

\subsection{Comparison of Time Series}
\begin{figure*}[htbp]
\begin{center}
\includegraphics[width=0.8\textwidth]{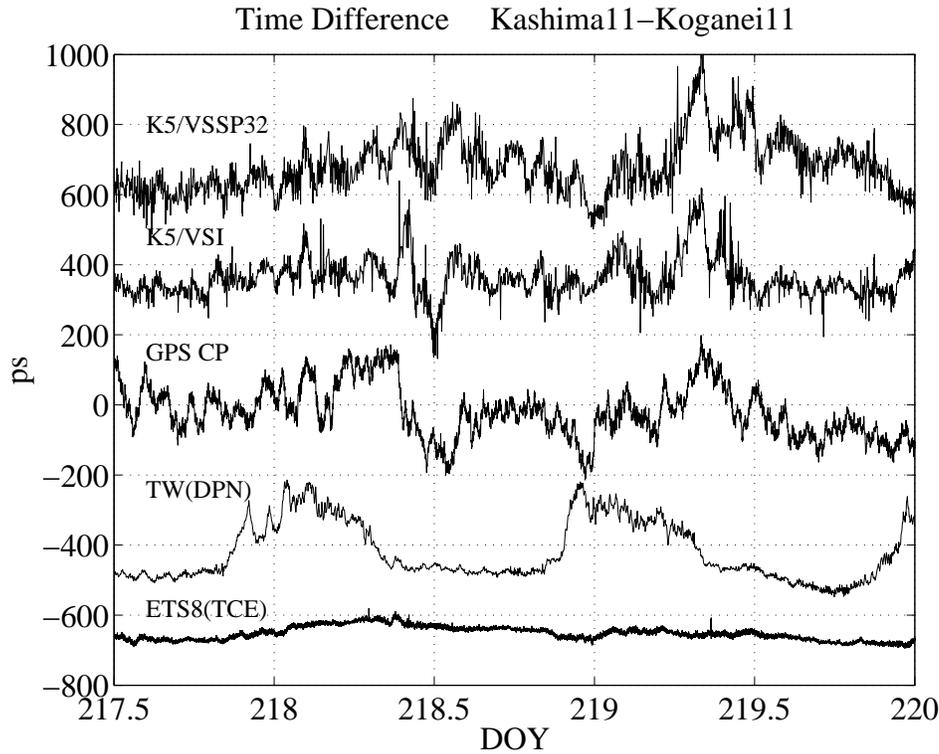}
\caption{Time difference obtained from these techniques.}
\label{times}
\end{center}
\end{figure*}

To evaluate long term stability of these techniques, we acquired the over 100 hours data at August experiment.
Figure \ref{times} show the time series of time difference calculated from these techniques at Kashima-Koganei baseline.
The common trend of these time series was already removed up to 2nd order.
Table \ref{integtime} show the data property (integration time, etc) of each techniques.
Also, Table \ref{rmsets8} show the root-mean-square of time series variation calculated from with reference to ETS8.
The result of ETS8(TCE) is extremely stable than other techniques.
TW(DPN) is also stable, but it has clear daily variation. The cause of the daily variation does not yet clear, but we think that it is caused by interference from spread signal and/or sunlight.
 
\begin{figure*}[htbp]
\begin{center}
\includegraphics[width=0.8\textwidth]{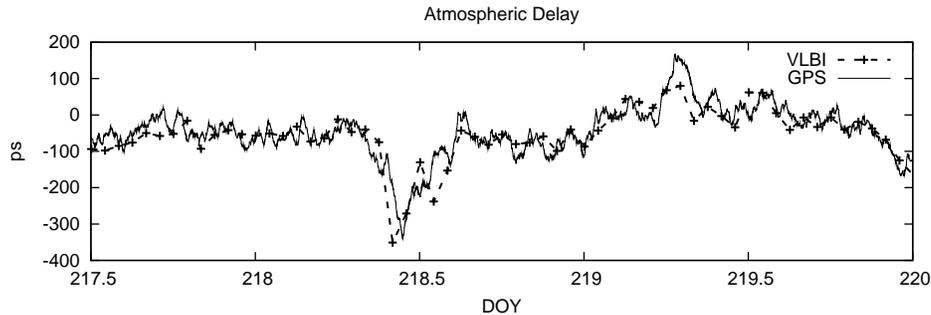}
\caption{The difference of the atmospheric delay calculated from VLBI and GPS between KSTC and Koganei Headquarters}
\label{atms}
\end{center}
\end{figure*}

The results of the VLBI agree with GPS, but these results vary than other results.
Figure \ref{atms} show the difference of the atmospheric delay calculated from VLBI and GPS between Kashima and Koganei.
Time delay and atmospheric delay variations agree well.
As already described, the influence of atmospheric delay was removed from TW(DPN) and ETS8(TCE) because both techniques are TWSTFT. 
Usually in the analysis of VLBI and GPS, the atmospheric delay and time delay are estimated at the same time.
These results suggested the estimation of atmospheric delay in the analysis of VLBI and GPS is not enough.
To obtain the more precise result, it is necessary to use the more precise model and/or another method such as KARAT \cite{hobiger2}.

\subsection{Comparison of T$\&$F transfer precision}
\begin{figure}[htdp]
 \begin{center}
  \includegraphics[width=70mm]{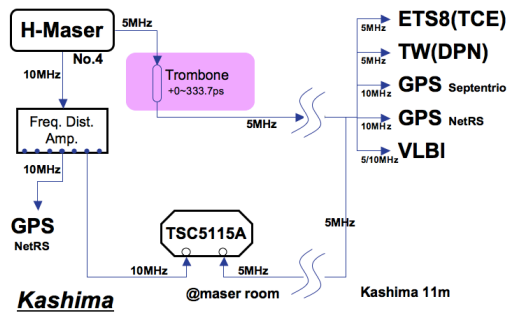}
  \caption{The reference signal setup diagram at Kashima station.}
  \label{keiro}             
\end{center}
\end{figure}

At October experiment, we compared the precision of these techniques by stretching the trombone which was inserted in the path of the reference signal from Hydrogen maser to the Kashima 11m antenna.
Figure \ref{keiro} show the reference signal setup diagram at Kashima station.
This experiment is almost same strategy in the case of \cite{taki4}.
In this time, we stretched the trombone more slowly and more constantly.
In addition, we expanded scan time of VLBI according to the time of stretching trombone.
Also, as the reference of correct change of time difference, we introduced the new DMTD equipment (TSC511A, Phase Noise and Allan Deviation Test Set).

\begin{figure*}[htbp]
\begin{center}
\includegraphics[width=0.9\textwidth]{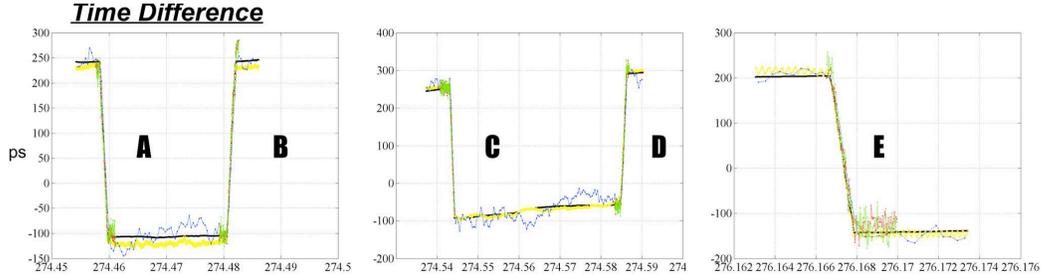}
\caption{Time difference of October experiment. The large steps (A to E) were artificial delay change parts by trombone. Black lines are DMTD. Gray thin lines are VLBI (VSSP and VSI) and ETS8(TCE). Variety lines are GPS.}
\label{steps}
\end{center}
\end{figure*}

Figure \ref{steps} show the time difference of each techniques. The large steps (A to E) were artificial delay change parts by trombone. Black lines are DMTD. Gray thin lines are VLBI (VSSP and VSI) and ETS8(TCE). Variety lines are GPS.
Also, we show the summary of the amount of the steps obtained from these techniques at the artificial delay change parts in Table \ref{trombone}.
These results show that each technique agree very well.
The differences of each technique except GPS are only a few picoseconds on the average.
Anyway, the result of our experiment clearly show that the geodetic VLBI technique can measure the correct time difference as same as ETS8(TCE) and DMTD.

 \section{Summary and Outlook}
We carried out the intercomparison experiments between VLBI and other techniques (GPS CP, TW(DPN), ETS8(TCE), and DMTD) to show the capability of VLBI time and frequency transfer by using the current geodetic VLBI technique and facilities as the summary of the experiments that we carried out since 2007.

The results from the August experiments show that the VLBI is more stable than GPS but is slightly noisier than two new two-way techniques (TW(DPN), ETS8(TCE)).
Also, these results show that the estimation of atmospheric delay in the analysis of VLBI and GPS is not enough.

At October experiment, we produced artificial delay changes by stretching the trombone which was inserted in the path of the reference signal from Hydrogen maser to Kashima 11m antenna.
At the artificial changes, the results of VLBI, ETS8 and DMTD hardly had a difference.
Consequently, the geodetic VLBI technique can measure the correct time difference as same as ETS8(TCE) and DMTD.

Currently the T$\&$F transfer experiment using ETS8 in NICT was finished.
And the project shifted to the next phase of the R$\&$D of T$\&$F transfer using Quasi-Zenith Satellite System (QZSS).
Also, the VLBI project shifted to the next phase which is the R$\&$D of the new facilities and strategy suitable for T$\&$F transfer.
In the near future, we are planning to carry out the following list.
\begin{itemize}
\item apply KARAT for the VLBI and GPS analysis
\item using MARBLE \cite{ishii} and ADS3000+ \cite{takefuji} for T$\&$F transfer
\item international experiment
\end{itemize}

\vspace{1.0\baselineskip}

\begin{table*}[htdp]
\caption{The setting parameters of experiments}
\begin{center}
\begin{tabular}{l|c|c}
 & K5/VSSP32 & K5/VSI (ADS1000) \\
 \hline \hline
Band & S/X & S/X \\
Input Freq. Width & 16MHz/ch & 512MHz/ch \\
Sampling Rate & 32Mbps, 1bit & 1024Mbps/ch, 1bit \\
Number of Channels & 16ch & 2ch \\
Effective Bandwidth of X-band & 364.8MHz & 147.8MHz \\
\hline
\end{tabular}
\end{center}
\label{vsspvsi}
\end{table*}

\begin{table}[htdp]
\caption{The baseline length calculated from the data that was sampled with each system of K5/VSSP32 and K5/VSI.}
\begin{center}
\begin{tabular}{l|c|c|c}
Exp. date & System & Baseline Length & 1$\sigma$ \\
\hline \hline
August & VSSP & 109099639.00 & 0.53 \\
            & VSI     & 109099639.00 & 0.57 \\
\hline
October & VSSP & 109099635.43 & 0.58 \\
              & VSI    & 109099635.58 & 0.66 \\
\hline
\multicolumn{4}{r}{Unit: mm} \\
\end{tabular}
\end{center}
\label{baseline}
\end{table}

\begin{table*}[htdp]
\caption{The data property of each techniques}
\begin{center}
\begin{tabular}{l|c|c|c|c|c}
 & \multicolumn{2}{c}{VLBI} & \multicolumn{3}{c}{} \\
 & VSSP & VSI & GPS CP & TW(DPN) & ETS8(TCE) \\
 \hline \hline
Integration Time & \multicolumn{2}{c}{Scan Length} & 1 & 120 & 1 \\
Analysis Software & K5 + Calc/Solve & GICO3 + Calc/Solve & NRCan PPP & \multicolumn{2}{c}{developed by NICT} \\
Data Interval & \multicolumn{2}{c}{Slew time + (Sum of two scan /2)} & 30 & 120 & 1 \\
& \multicolumn{2}{c}{Average 100} & & & \\
\hline
\multicolumn{5}{r}{Unit: second} \\
\end{tabular}
\end{center}
\label{integtime}
\end{table*}

\begin{table}[htdp]
\caption{The root-mean-square of time series variation calculated from with reference to ETS8.}
\begin{center}
\begin{tabular}{l|c|c}
 & August & October \\
\hline \hline
TW(DPN) & 95 (20) & - \\
GPS CP & 75 & 56 \\
VLBI(VSI) & 60 & 36 \\
\hline
\multicolumn{3}{r}{Unit: ps} \\
\end{tabular}
\end{center}
\label{rmsets8}
\end{table}

\begin{table}[htdp]
\caption{The amount of the steps obtained from these techniques at the artificial delay change parts.}
\begin{center}
\begin{tabular}{l|c|c|c|c|c|c}
 & A & B & C & D & E & Average \\
\hline \hline
DMTD         & 347 & 346 & 346 & 347 & 348 & 347 \\
GPS CP      & 352 & 340 & 385 & 353 & 345 & 355 \\
ETS8  & 348 & 340 & 343 & 349 & 347& 345 \\
VSSP & -      & 349 & 342 & 350 & 340 & 345 \\
VSI    & 347 & 340 & -      & 351 & 348 & 346 \\
\hline
\multicolumn{7}{r}{Unit: ps} \\
\end{tabular}
\end{center}
\label{trombone}
\end{table}

\hspace{-1em}{\it Acknowledgments}:
The authors would like to acknowledge the IVS and the IGS for the high quality products. We are grateful that GSFC and NR Canada provided the VLBI and GPS analysis software (CALC/SOLVE and NRCan's PPP).
 The VLBI experiments were supported by M. Sekido and E. Kawai of the Kashima Space Technology Center.


\begin{thebibliography}{9}

\fontsize{6pt}{0pt}\selectfont

  \bibitem{taki1}
Takiguchi, H., T. Hobiger, A. Ishii, R. Ichikawa, and Y. Koyama, Comparison with GPS Time Transfer and VLBI Time Transfer, {\it IVS NICT-TDC News}, No.28, 10-15, 2007.

 \bibitem{taki2}
Takiguchi, H., Y. Koyama, R. Ichikawa, T. Gotoh, A. Ishii, T. Hobiger and M. Hosokawa, Comparison Study of VLBI and GPS Carrier Phase Frequency Transfer using IVS and IGS data, {\it IVS NICT-TDC News}, No.29, 23-27, 2008.

 \bibitem{taki3}
Takiguchi, H., Y. Koyama, R. Ichikawa, T. Gotoh, A. Ishii, and T. Hobiger,  Comparison Study of VLBI and GPS Carrier Phase Frequency Transfer - Part II -, {\it IVS NICT-TDC News}, No.30, 26-29, 2009.

 \bibitem{taki4}
Takiguchi, H., Y. Koyama, R. Ichikawa, T. Gotoh, A. Ishii, T. Hobiger, and M. Hosokawa, VLBI Measurements for Frequency Transfer, {\it IVS NICT-TDC News}, No. 31, 21-24, 2010.

\bibitem{amagai}
Amagai, J., and T. Gotoh, Current Status of Two-way Satellite Time and Frequency Transfer Using a Pair of Pseudo Random Noises, {\it Proc. ATF 2008}, 2008.

\bibitem{takahashi1}
Takahashi, Y., M. Imae, T. Gotoh, F. Nakagawa, H. Kiuchi, M. Hosokawa, M. Aida, Y. Takahashi, H. Noda, and S. Hama, 
Time Comparison Equipment for ETS-VIII Satellite -Part 1 Development of Flight Model-, 
{\it J. NICT}, Vol.50, Nos.1/2, 4-4, 135-143, 2003.

\bibitem{fujieda1}
Fujieda, M., Y. Takahashi, T. Gotoh, F. Nakagawa, and M. Imae, 
Time Comparison Experiment, Earth Station of Time Comparison Experiment, 
{\it J. NICT}, Vol.50, Nos.3/4, 4-8-1, 243-249, 2003.

\bibitem{nakagawa}
Nakagawa, F., M. Imae, Y. Takahashi, H. Kiuchi, T. Gotoh, and M. Fujieda,
Time Comparison Experiment, Time Comparison Equipment for ETS-VIII -Data Processing, Analysis and Capability of Time Comparison-, 
{\it J. NICT}, Vol.50, Nos.3/4, 4-8-2, 251-259, 2003.


 \bibitem{fujieda2}
Fujieda, M., M. Kumagai, S. Nagano, and T. Gotoh, UTC(NICT) signal transfer system using optical fibers, {\it IVS NICT-TDC News}, No. 31, 17-20, 2010.

 \bibitem{koyama}
Koyama, Y., T. Kondo, M. Sekido, and M. Kimura, Developments of K5/VSI System for Geodetic VLBI Observations, {\it IVS NICT-TDC News}, No. 29, 15-20, 2008. 

\bibitem{kondo}
Kondo, T., Y. Koyama, R. Ichikawa, M. Sekido, E. Kawai, and M. Kimura, 
Development of the K5/VSSP System, 
{\it J. Geodetic Soc. Japan}, 54(4), 233-248, 2008.

\bibitem{kimura}
Kimura, M., J. Nakajima, T. Kondo, Y. Koyama, M. Sekido, and T. Oyama, 
High Speed VLBI Software Correlator {\it (now preparing)}.

\bibitem{hobiger1}
Hobiger, T., R. Ichikawa, and Y. Koyama, MK3TOOLS - seamless interfaces for the creation of VLBI databases from post-correlation output, {\it Proc. the Fifth IVS General Meeting}, 153–156, 2008.

\bibitem{hobiger2}
Hobiger, T., R. Ichikawa, Y. Koyama, and T. Kondo, Kashima ray-tracing service (KARATS) online provision of total troposphere slant delay corrections for east asian sites, {\it Proc. International Symposium on GPS/GNSS 2008}, 1, 40–44, 2008.

\bibitem{takefuji}
Takefuji, K., M. Tsutsumi, H. Takeuchi, and Y. Koyama, 
Current Status of Next Generation A/D Sampler ADS3000+, 
{\it IVS NICT-TDC News}, No. 31, 6-9, 2010.

\bibitem{ishii}
Ishii, A., R. Ichikawa, H. Takiguchi, K. Takefuji, H. Ujihara, Y. Koyama, T. Kondo, S. Kurihara, Y. Miura, S. Matsuzaka and D. Tanimoto, 
Current status of development of a transportable and compact VLBI system by NICT and GSI, 
{\it IVS NICT-TDC News}, No. 31, 2-5, 2010.



\end{thebibliography}
\end{document}